\documentclass[a4paper,12pt, epsfig]{article}
\def\letter{0}\def\pr{0}
\usepackage{epsfig}
\usepackage{epstopdf}
\usepackage{graphicx}
\usepackage{ifthen}

\usepackage{amsmath}
\usepackage[psamsfonts]{amssymb}
\usepackage{euscript}

\usepackage{latexsym}
\usepackage[arrow,matrix,curve]{xy}

\newskip\humongous \humongous=0pt plus 1000pt minus 1000pt

\newif\ifdtup

\def\,{\hspace{-.1cm}}
\def\hsp{,\hspace{.7cm}}

\def\tf {\tilde{f}}
\def\ff{{\mathcal{F}}}

\def\fc#1#2 {\frac{n}{q}#1\frac{n}{q}#2}

\def\hf{H\p_{f_0}}

\newcommand{\vac}{\ensuremath{|0\rangle}}
\newcommand{\stt}{\ensuremath{|\psi\rangle}}

\renewcommand{\cos}{\textrm{cos}}
\renewcommand{\sin}{\textrm{sin}}

\renewcommand{\sinh}{\textrm{sinh}}
\renewcommand{\cosh}{\textrm{cosh}}
\renewcommand{\tanh}{\textrm{tanh}}
\newcommand{\sech}{\textrm{sech}}
\newcommand{\csch}{\textrm{csch}}

\def\exp#1{\hbox{\rm exp}\left(#1\right)}

\renewcommand{\theequation}{\arabic{section}.\arabic{equation}}
\renewcommand{\(}{\begin{equation}}
\renewcommand{\)}{end{equation} \vspace{-.05in}\linebreak}

\newcounter{saveeqn}
\newcounter{savealpheqn}

\newcommand{\alpheqn}{\setcounter{saveeqn}{\value{equation}}%
  \stepcounter{saveeqn}\setcounter{equation}{0}%
  \renewcommand{\theequation}{\mbox{\arabic{section}.\arabic{saveeqn}
\alph{equation}}}
  \renewcommand{\)}{\end{equation}}}
\def\part#1{\frac{\partial}{\partial{#1}}}%
\def\group#1{\refstepcounter{equation}\setcounter{saveeqn}
 {\value{equation}}%
  \label{#1}\setcounter{equation}{0}%
\renewcommand{\theequation}{\mbox{\arabic{section}.\arabic{saveeqn}
\alph{equation}}}
  \renewcommand{\)}{\end{equation}}}
\newcommand{\reseteqn}{\setcounter{equation}{\value{saveeqn}}%
  \renewcommand{\theequation}{\arabic{section}.\arabic{equation}}%
  \renewcommand{\)}{\end{equation}}}

\newcommand{\aalpheqn}{\setcounter{saveeqn}{\value{equation}}%
  \stepcounter{saveeqn}\setcounter{equation}{0}%
  \renewcommand{\theequation}{\mbox{
        \Alph{subsection}.\arabic{saveeqn}\alph{equation}}}
   \renewcommand{\)}{\end{equation}}}
\newcommand{\areseteqn}{\setcounter{equation}{\value{saveeqn}}%
  \renewcommand{\theequation}{\Alph{subsection}.\arabic{equation}}%
  \renewcommand{\)}{\end{equation}}}

\renewcommand{\thefootnote}{\alph{footnote}}
\renewcommand{\(}{\begin{equation}}
\renewcommand{\)}{\end{equation}}
\newcommand{\ba}{\begin{eqnarray}}
\newcommand{\ea}{\end{eqnarray}}

\renewcommand{\b}{\beta}

\newcommand{\cbp}{\mathop{\vtop{\ialign{##\crcr
   $\hfil\displaystyle{}\hfil$\crcr\noalign{\kern-13pt\nointerlineskip}
   \BIG{)}\hskip0pt\crcr\noalign{\kern3pt}}}}}
\newcommand{\pa}{\mathop{\vtop{\ialign{##\crcr

$\hfil\displaystyle{\oplus}\hfil$\crcr\noalign{\kern+1pt\nointerlineskip
}
   \hspace{.08in}$^{\alpha=0}$\hskip6pt\crcr\noalign{\kern3pt}}}}}
\renewcommand{\hsp}{,\hspace{.3in}}
\newcommand{\p}{^\prime}
\newcommand{\pp}{^{\prime\prime}}

\def\D{\ensuremath{{\cal D}}}

\catcode`\@=11
\def\vereq#1#2{\lower3pt\vbox{\baselineskip1.5pt \lineskip1.5pt
\ialign{$\m@th#1\hfill##\hfil$\crcr#2\crcr\sim\crcr}}}
\catcode`\@=12

\renewcommand{\(}{\begin{equation}}
\renewcommand{\)}{\end{equation}}

\def\pin#1{\int \frac{d#1}{2\pi}}
\def\ppin#1{\int\hspace{-17pt}\sum \frac{d#1}{2\pi}}
\def\ppink#1{\int\hspace{-17pt}\sum\frac{d^{#1}k}{(2\pi)^{#1}}}
\def\dint{\int\hspace{-12pt}\sum }
\def\pink#1{\int \frac{d^{#1}k}{(2\pi)^{#1}}}

\def\Bd#1{B^\dag_{k_{#1}}}

\def\tp#1#2#3{\hbox{\rm tan}^#1(\thet#2#3)}

\def\cc{\mathcal{C}}
\def\df{\mathcal{D}_{f}}

\def\dF{\mathcal{D}_F}

\def\I{\mathcal{I}}

\def\os{\omega_S}

\def\as{|\alpha;\sigma\rangle}
\def\asb{\langle\alpha;\sigma|}

\newcommand{\beas}{\begin{eqnarray*}}
\newcommand{\eeas}{\end{eqnarray*}}

\newcommand{\bquo}{\begin{quote}}
\newcommand{\enqu}{\end{quote}}

\newcommand{\g}{{\mathfrak g}}

\def\ch{{\mathcal{H}}}

\def\tp{{\tilde{\phi}}}
\def\ok#1{\omega_{k_{#1}}}

\def\V#1{V^{(#1)}(gf(x))}

\newcommand{\beq}{\begin{equation}}
\newcommand{\eeq}{\end{equation}}
\newcommand{\bea}{\begin{eqnarray}}
\newcommand{\eea}{\end{eqnarray}}

\newskip\humongous \humongous=0pt plus 1000pt minus 1000pt

\newif\ifdtup

\catcode`\@=11

\ifthenelse{\equal{\letter}{0}}{ 

\@addtoreset{equation}{section}
\def\theequation{\arabic{section}.\arabic{equation}}

\def\@normalsize{\@setsize\normalsize{15pt}\xiipt\@xiipt
\abovedisplayskip 14pt plus3pt minus3pt%
\belowdisplayskip \abovedisplayskip
\abovedisplayshortskip \z@ plus3pt%
\belowdisplayshortskip 7pt plus3.5pt minus0pt}

\def\small{\@setsize\small{13.6pt}\xipt\@xipt
\abovedisplayskip 13pt plus3pt minus3pt%
\belowdisplayskip \abovedisplayskip
\abovedisplayshortskip \z@ plus3pt%
\belowdisplayshortskip 7pt plus3.5pt minus0pt
\def\@listi{\parsep 4.5pt plus 2pt minus 1pt
      \itemsep \parsep
      \topsep 9pt plus 3pt minus 3pt}}

\relax

\def\section{\@startsection{section}{1}{\z@}{3.5ex plus 1ex minus  .2ex}{2.3ex plus .2ex}{\large\bf}}

\def\thesection{\arabic{section}}
\def\thesubsection{\arabic{section}.\arabic{subsection}}

\def\appendix{\setcounter{section}{0}
 \def\thesection{Appendix \Alph{section}}
 \def\thesubsection{\Alph{section}.\arabic{subsection}}
 \def\theequation{\Alph{section}.\arabic{equation}}}
\renewcommand{\theequation}{\arabic{section}.\arabic{equation}}

}{
\renewcommand{\theequation}{\arabic{equation}}

} 

\begin{document}
\def\thefootnote{\fnsymbol{footnote}}
\def\thetitle{Form Factors for Meson-Kink Scattering}
\def\autone{Jarah Evslin}
\def\auttwo{Hengyuan Guo}
\def\affa{Institute of Modern Physics, NanChangLu 509, Lanzhou 730000, China}
\def\affb{University of the Chinese Academy of Sciences, YuQuanLu 19A, Beijing 100049, China}

\title{Titolo}

\ifthenelse{\equal{\pr}{1}}{
\title{\thetitle}
\author{\autone}
\affiliation {\affa}
\affiliation {\affb}

}{

\begin{center}
{\large {\bf \thetitle}}

\bigskip

\bigskip


{\large \noindent  \autone{${}^{1,2}$} }


\vskip.7cm

1) \affa\\
2) \affb\\

\end{center}

}

\begin{abstract}
\noindent
We calculate the leading quantum corrections to the meson form factors of nonrelativistic kinks, at momentum transfer much higher than the meson mass.  We consider general scalar theories which need not be integrable.  Our approach is much simpler than previous approaches, avoiding any need for counterterms, ordering ambiguities and nonlinear canonical transformations at any order in perturbation theory.


\end{abstract}

%
\setcounter{footnote}{0}
\renewcommand{\thefootnote}{\arabic{footnote}}

\ifthenelse{\equal{\pr}{1}}
{
\maketitle
}{}

\section{Introduction}

In the limit of many colors, baryons in (3+1)-dimensional quantum chromodynamics (QCD) become solitons constructed from a scalar meson field~\cite{skyrme,wittenskyrme}.  Investigating the classical properties of this system is clearly of interest and there has been steady progress for decades~\cite{zahedskyrme,smorg}.  The corresponding quantum system, while essential for understanding QCD, is difficult to study due to the large number of degrees of freedom \cite{christesi}.  This has motivated the study of (1+1)-dimensional toy models of a scalar meson field together with a kink soliton solution \cite{dhn2,rajaraman75,weigrev}.  The hope has been that such models provide a laboratory in which one may construct methods \cite{slavatrunc,serone} for treating the quantum system which may eventually be applied to the (3+1)-dimensional theory realized in Nature.  Such methods need to be simple in (1+1) dimensions if they are to be of use in (3+1) dimensions.

In these (1+1)-dimensional toy models, meson-baryon scattering is replaced by meson-kink scattering.  While meson-baryon scattering is an important topic in QCD with a rich phenomenology, as well as important implications for meson-mediated baryon-baryon interactions, quantum meson-kink scattering has so far received only modest attention \cite{swanson88,uehara91,abdel11} beyond integrable models \cite{zam77}.  

In this letter we solve the critical problem required by any study of meson-kink scattering, we provide a simple method to calculate the form factors describing the absorption or emission of an ultrarelativistic meson by a nonrelativistic kink.  Using this method, we calculate the leading quantum corrections to the form factors.  Further details regarding the calculations and consistency checks can be found in Ref.~\cite{meff}.

\section{The Models}

Consider a (1+1)-dimensional theory of a mass $m$ Schrodinger picture scalar field $\phi(x)$ and its conjugate $\pi(x)$, described by the Hamiltonian
\beq
H=\int dx :\left[  \frac{\pi^2(x)}{2}+\frac{\left(\partial_x\phi(x)\right)^2}{2}+\frac{V(g\phi(x))}{g^2}\right]:_a\nonumber
\eeq
where $::_a$ is the usual normal ordering and $g$ is a small coupling constant.  If the potential $V$ has multiple minima, then there will be kink solutions $f(x)$ of the classical equations of motion
\bea
\phi(x,t)=f(x),\  &&
f\pp(x)=\frac{1}{g}\V{1}\\
\V{n}&=&\frac{\partial^n}{\partial(g\phi(x))^n}V(g\phi(x))|_{\phi(x)=f(x)}.\nonumber
\eea

Quantum states corresponding to the kink and its excitations lie in the kink sector of the Hilbert space, which is related to the vacuum sector by the unitary displacement operator
\beq
\df={\rm{exp}}\left(-i\int dx f(x)\pi(x)\right). \label{df}
\eeq
Acting the operator $\df$ on a vacuum sector state produces a kink sector state.  Let $|K\rangle$ be a Hamiltonian eigenstate in the kink sector.  For concreteness we will choose the ground state, but the extension to other Hamiltonian eigenstates is trivial.  Then
\beq
\vac=\df^\dag|K\rangle
\eeq
is a vacuum sector state.  As $|K\rangle$ is an eigenvector of the Hamiltonian $H$,  $\vac$ is an eigenvector of the kink Hamiltonian $H\p$
\beq
H\p=\df^\dag H\df,\ 
H\df\vac=Q\df\vac\Rightarrow
H\p\vac=Q\vac.
\eeq
Similarly, the state $\vac$ may be boosted using $\Lambda\p=\df^\dag \Lambda\df$ where $\Lambda$ is the usual boost operator.

This formalism is useful because $\vac$ can be found by solving the $H\p$ eigenvalue problem in perturbation theory \cite{mekink,me2loop}, decomposing in powers of $g$
\beq
\vac=\sum_{i=0}^\infty \vac_i,\  Q=\sum_{i=0}^\infty Q_i,\ 
\Lambda\p=\sum_{i=1}^\infty \Lambda\p_i.
\eeq
First, one decomposes $H\p$ and $\Lambda\p$ it into pieces $H_i$ and $\Lambda\p_i$ with products of $i$ fields when normal ordered.  Now $H_0$ is a scalar, the classical kink mass $Q_0$, while $H_1$ vanishes.  

The first nontrivial operator is $H_2$, which resembles a free Hamiltonian but with a position-dependent mass term.  It may be diagonalized by decomposing the fields into an orthonormal basis of kink normal modes.  The normal modes are the real zero mode
\beq
\g_B(x)=f\p(x)/\sqrt{Q_0} \label{gb}
\eeq
with frequency $\omega_B=0$, the complex continuum modes $\g_k(x)$ for all real $k$ with frequency $\omega_k=\sqrt{m^2+k^2}$ and also, sometimes, real, discrete shape modes $\g_S(x)$ with frequency $\omega_S<m$.   Then we define the decomposition
\beq
\tp_k=\int dx \phi(x) \g^*_k(x)\hsp
\tilde{\pi}_k=\int dx \pi(x) \g^*_k(x)
\eeq
where $k$ runs over $B$, $S$ and all real values.  We will write $\phi_0$ and $\pi_0$ instead of $\tp_B$ and $\tilde{\pi}_B$ below.  In the case of the discrete and continuous nonzero modes, these can be reorganized in terms of creation and annihilation operators
\beq
B^\dag_k=\frac{\tp_k}{2}-i\frac{\tilde{\pi}_k}{2\omega_k}\hsp
\frac{B_{-k}}{2\omega_k}=\frac{\tp_k}{2}+i\frac{\tilde{\pi}_k}{2\omega_k}.
\eeq
This decomposition provides a new basis $\phi_0$, $\pi_0$, $B^\dag$ and $B$ of the algebra of operators, satisfying canonical commutation relations and the oscillator algebra respectively.  This decomposition in the kink sector plays a role similar to the Fourier transformed basis
\beq
\tp_p=\int dx \phi(x)e^{ipx}\hsp
\tilde{\pi}_p=\int dx \pi(x)e^{ipx} \label{pdec}
\eeq
in the vacuum sector.


In terms of the normal mode basis, the Hamiltonian term $H_2$ is simply
\beq 
H_2=Q_1+\frac{\pi_0^2}{2}+\os B^\dag_SB_S+\ppin{k}\ok{} B^\dag_kB_k \label{h2p}
\eeq
in the case of a single shape mode.  The first term, $Q_1$, is the one-loop kink mass.  The second is the kinetic energy of the kink center of mass and the last is a sum of quantum harmonic oscillators.  The notation  $\dint$ denotes an integral over all real $k$ plus a sum over shape modes.

The spectrum of $H_2$ is easy to write down.  In particular, the ground state $\vac_0$ is the unique state that satisfies
\beq
\pi_0\vac_0=B_S\vac_0=B_k\vac_0=0.
\eeq
Excited states are created by acting with $B^\dag_k$ and $B_S^\dag$.


\section{Form Factors}

Define the position eigenstate kink states $|y\rangle_0$ by
\beq
\phi_0|y\rangle_0=y|y\rangle_0\hsp
B_k|y\rangle_0=B_S|y\rangle_0=0
\eeq
where $y/\sqrt{Q_0}$ is the position of the kink.  Then our leading order ground state is, up to an arbitrary normalization
\beq
\vac_0=\int dy |y\rangle. \label{y}
\eeq
One sees that the Hamiltonian eigenstates are necessarily nonnormalizable, as they contain sums over all possible kink centers of mass on the real line.  A more serious problem is that our perturbative approach does not converge, even in the sense of an asymptotic series, at $|y|>1/\sqrt{mg}$, which occupies most of the wave function of $\vac_0$.  

This motivates us to introduce wave packets normalized to unity via the constant $\mathcal{N}$
\beq
\as=\frac{\sqrt{\mathcal{N}}e^{-\frac{\phi_0^2}{4\sigma^2}}}{(2\pi)^{1/4}\sqrt{\sigma}}e^{i\alpha\Lambda\p}\vac,\ 
\langle\alpha;\sigma\as=1. \label{norm}
\eeq
These wave packets, or more precisely the states $\df\as$, have a position-space width of $\sigma/\sqrt{Q_0}$ and expected rapidity $\alpha$.   We will be interested in the form factors relevant to the scattering of such localized wave packets of kinks.  However at the end of this letter we suggest that our results also yield the form factors of delocalized, momentum eigenstate kinks in the case of ultrarelativistic mesons.

The form factor $\tilde{\ff}_q$ and its transform $\ff(z)$ are
\beq
\tilde{\ff}_q=\langle 0;\sigma|\df^\dag \tp_q \df\as=\int dz \ff(z) e^{iqz}. \label{ffdef}
\eeq
Note that the wave packet has a distribution of momenta while the operator $\tp_q$ has a fixed momentum.  Translation-invariance implies momentum conservation and so one expects the form factor to be maximized at $q=-Q_0 \alpha$, where the maxima of the momenta of the wave functions of the bra and the ket differ by $q$ units.

Now we will calculate the form factor at leading order.  At leading order, the wave packet is \cite{meimpulso}
\beq
\as_0=\frac{e^{-\frac{\phi_0^2}{4\sigma^2}}e^{i\alpha\Lambda\p_1}}{(2\pi)^{1/4}\sqrt{\sigma}}\vac_0,\ \Lambda\p_1=-\sqrt{Q_0}\phi_0 \label{sig0}
\eeq
and so the leading order form factor is
\beq
\tilde{\ff}_{{0},q}={}_0\langle 0;\sigma|\df^\dag \tp_q \df \as_0.
\eeq
Using the property
\beq
\df^\dag \phi(x) \df=\phi(x)+f(x)
\eeq
together with Eqs.~(\ref{gb}) and (\ref{y}) we find
\bea
\tilde{\ff}_{{0},q}&=&\frac{1}{\sigma\sqrt{2\pi}}\int dx e^{iqx}\int dy e^{-\frac{y^2}{2\sigma^2}-i\alpha\sqrt{Q_0}y}\nonumber\\
&&\times\left(f(x)+\frac{y}{\sqrt{Q_0}}f\p(x)
\right). \nonumber
\eea
Now define the coordinate $z$ relative to the kink by
\beq
z=x+\frac{y}{\sqrt{Q_0}}
\eeq
and the offset $\epsilon$ of the momentum peak of the wave function from the value that is imposed by momentum conservation
\beq
\epsilon=Q_0\alpha+q. \label{ep}
\eeq
Then, fixing $\epsilon$, the Fourier transform of the above expression is
\bea
&&\ff_0(z)\,=\,\int\, dy \frac{e^{-\frac{y^2}{2\sigma^2}-i\epsilon y/\sqrt{Q_0}}}{\sigma\sqrt{2\pi}}\,\left(\,f(z)-\frac{y^2}{2Q_0}f^{\prime\prime}(z)\,\right)\nonumber\\
&&\ =e^{-\frac{\sigma^2\epsilon^2}{2Q_0}}\,\,\left[f(z)-\frac{\sigma^2}{2Q_0}\left(1-\frac{\sigma^2\epsilon^2}{Q_0}\right)f\pp(z)\right] \label{ff0}
\eea
up to corrections of order $Q_0^{-3/2}$.  

At $\epsilon=0$ the first term is the classical result \cite{gj75} that the Fourier transformed form factor is the classical solution.  For nonzero $\epsilon$, there is a suppression factor which unsurprisingly resembles the momentum space wave function.  The second term is suppressed by a power of $Q_0^{-1}$.  Thus the first term is of order $O(1/g)$ and the second of order $O(g)$.

We have found the first quantum correction to the form factor.  There are three other sources of quantum corrections.  First, there are corrections from the higher order boost operator.  We have calculated the corrections resulting from $\Lambda\p_2$ and $\Lambda\p_3$ and found that both appear at order $O(g^3)$, and so they will not be discussed here.  Next, there are corrections resulting from the normalization.  There is a one-to-one correspondence between those corrections, and corrections resulting from the matrix elements of the scalar $f(x)$ between the corrections $\vac_{i>0}$ to the kink ground state.  At $\epsilon=0$ these corrections cancel, at least at order $O(g)$, while more generally a calculation similar to that above yields, at $O(g)$, a total correction of
\beq
\,\,\,\cc_{\mathcal{N}}(z)\,=\,f(z)\frac{\sigma^4\epsilon^2}{Q_0^2}e^{-\frac{\sigma^2\epsilon^2}{2Q_0}}\,\left[\,-2 M_{02}+\,\left(\,-6\,+\,\frac{\sigma^2\epsilon^2}{Q_0}\,\right)\, M_{22}\,\right]\label{nc}
\eeq
where 
\bea
M_{ij}&=&\,\ppin{k}\frac{\gamma_1^{i1}(-k)\gamma_1^{j1}(k)}{2\ok{}}\\
\vac_1\,&=&\,\,\frac{1}{\sqrt{Q_0}}\ppin{k}\left[\gamma_1^{01}(k)+\phi_0^2 \gamma_1^{21}(k)\right]B^\dag_k\vac_0.\nonumber
\eea

The most interesting correction to the form factor at order $O(g)$ results from the corrections $\vac_1$ and the corresponding matrix elements of the operator $\phi(x)$.   Consider the subleading wave packet term
\beq
\as_{1}=\frac{1}{(2\pi)^{1/4}\sqrt{\sigma}}e^{-\frac{\phi_0^2}{4\sigma^2}}e^{i\alpha\Lambda\p_1}\vac_1.
\eeq
The leading correction to the form factor is
\beq
{}_0\langle 0;\sigma|\df^\dag \tp_q \df \as_{1}+{}_{1}\langle 0;\sigma|\df^\dag \tp_q \df \as_{0}\nonumber\\
\eeq
whose Fourier transform at fixed $\epsilon$ is
\bea
&&\frac{2}{\sqrt{Q_0}}e^{-\frac{\sigma^2\epsilon^2}{2Q_0}}
\ppin{k} \frac{\g_{-k}\left(z\right)}{2\ok{}} \\
&&\times\left[{\gamma_1^{01}(k)}+\sigma^2\left(1-\frac{\sigma^2\epsilon^2}{Q_0}\right){\gamma_1^{21}(k)}
\right].\nonumber
\eea
Using the known \cite{me2loop} form of $\gamma_1^{21}$
\beq
\gamma_1^{21}(k)=\frac{\ok{}\Delta_{kB}}{2}
\eeq
one sees that the $\gamma_1^{21}$ term cancels precisely with the quantum correction found in Eq.~(\ref{ff0}), leaving only the $\gamma_1^{01}$ term.

This term can be evaluated using \cite{me2loop}
\bea
\gamma_1^{01}(k)\,&=&\,\frac{\Delta_{kB}}{2}\,-\,\frac{g\sqrt{Q_0}}{2\ok{}}\int dx \V3 \I(x) \g_k(x) \nonumber\\
\I(x)&=&\pin{k}\frac{\left|{\g}_{k}(x)\right|^2-1}{2\omega_k}+\sum_S \frac{\left|{\g}_{S}(x)\right|^2}{2\omega_k}.\label{di}
\eea
In all, one finds that this correction is
\bea
\cc(z)
&=&\frac{1}{\sqrt{Q_0}}e^{-\frac{\sigma^2\epsilon^2}{2Q_0}}
\int dx\ppin{k} \frac{\g_{-k}\left(z\right)\g_k(x)}{2\ok{}}\nonumber\\
&&\, \left[\g_B\p(x)-\frac{g\sqrt{Q_0}\V3\I(x)}{\ok{}}
\right].
\eea
This is the only quantum correction to the form factor at order $O(g)$ when $\epsilon=0$, whereas for general $\epsilon$ there is also a normalization correction (\ref{nc}).

\section{Delocalized Kinks}

We have checked \cite{meff} that, at leading order in $q/m$, and at $\sigma=0$ these corrections agree with the leading corrections to the Sine-Gordon and $\phi^4$ form factors calculated in Refs.~\cite{weisz77} and \cite{gervaisff75} respectively.  Those papers did not consider the narrow kink wave packets considered here, but rather exact energy eigenstates.  Naively this would correspond to the case $\sigma=\infty$, where values of $|y|$ become much larger than $1/\sqrt{mg}$ and so one expects our perturbative approach to fail.  The integrability and collective coordinate approaches considered in those papers do not expand perturbatively in $\phi_0$ or its eigenvalue $y$ and so do not share our limitation.

We believe the reason for this agreement is as follows.  While we are limited to small values of the kink center of mass position $|y|$ by our perturbative approach, the momentum eigenstates considered in Refs.~\cite{weisz77,gervaisff75} are translation-invariant.  For each coordinate $z$ in the kink frame, translation-invariance may be used to set $y=0$.  Thus the $y=0$ terms in our expansions should capture the $\sigma=\infty$ behavior.  We always expand the form factors in a power series in $y$, with nonconstant terms giving powers of $\sigma$.  Thus, at $\sigma=0$, one arrives at the answer at $y=0$, where our perturbative expansion in $y$ truncates at the zeroeth term and so does not lead to any error.  

There is however some error introduced by the $e^{-\phi_0^2/4\sigma^2}$ term in our wave packet, which smears the momentum.  This smearing is larger than the meson mass $m$.  Thus our wave packet form factors will be smeared with respect to form factors for momentum eigenstates.  This smearing is larger than $m$.  As a result, we expect agreement of the $\sigma=0$ part of our form factors with the form factors for exact momentum eigenstates only at leading order in $q/m$, corresponding to scattering with ultrarelativistic mesons.

We note that while our approach is restricted to nonrelativistic kinks, as a result of our perturbative expansion, there is no such restriction to the approach to form factors in Ref.~\cite{andyff2}.

\section* {Acknowledgement}

\noindent
JE is supported by the CAS Key Research Program of Frontier Sciences grant QYZDY-SSW-SLH006 and the NSFC MianShang grants 11875296 and 11675223.   JE also thanks the Recruitment Program of High-end Foreign Experts for support.

\end{document}